\documentclass[authoryear,preprint,review,12pt]{elsarticle}

\usepackage{longtable} 
\usepackage{supertabular}
\usepackage{graphicx}
\usepackage{amssymb}
\usepackage{amsmath}
\usepackage{units}
\usepackage{version}
\usepackage{color}
\usepackage{float}

\usepackage{amssymb}

\reversemarginpar

\def\be{\begin{equation}}
\def\ee{\end{equation}}
\def\bi{\begin{itemize}}
\def\ei{\end{itemize}}
\newcommand{\Archeops}{{\sc Archeops}}
\newcommand{\Wmap}{{\sc Wmap}}
\newcommand{\Planck}{{\sc Planck}}
\newcommand{\HEALpix}{{\sc HEALpix}}
\newcommand{\Hammurabi}{{\sc Hammurabi}}

\journal{Astroparticle Physics}

\begin{document}

\begin{frontmatter}



\title{Model of the polarized foreground diffuse Galactic emissions from 33 to 353 GHz}


\author{L.~Fauvet, J.~F.~Mac\'{\i}as-P\'erez}

\address{LPSC, Universit\'e Joseph Fourier Grenoble 1, CNRS/IN2P3,
  Institut National Polytechnique de Grenoble, 53 avenue des Martyrs,
  38026 Grenoble cedex, France}

\author{F.X. D\'esert}

\address{IPAG: Institut de Plan\'etologie et
d'Astrophysique de Grenoble, UJF-Grenoble 1 / CNRS-INSU, UMR 5274,
Grenoble, F-38041, France}

\begin{abstract}

We present 3D models of the Galactic magnetic field including regular and
turbulent components, and of the distribution of matter in the Galaxy
including relativistic electrons and dust grains.
By integrating along the line of sight, we construct maps of the
polarized Galactic synchrotron and
thermal dust emissions for each of these models. We perform a
likelihood analysis to compare the maps of the Ka, Q, V and W bands of the Wilkinson
Microwave
Anisotropy Probe (\Wmap) and the 353 GHz \Archeops\ data to the models
obtained by varying the
pitch angle of the regular magnetic field, the relative amplitude of
the turbulent magnetic field and the extrapolation spectral
indices of the synchrotron and thermal dust emissions. The best-fit parameters obtained for the different
frequency bands are very similar and globally the data seem to favor a negligible
isotropic turbulent magnetic field component at large angular scales (an anisotropic
line-of-sight ordered component can not be studied using these data). From this study, we conclude that we are able to propose a consistent model
  of the polarized diffuse Galactic synchrotron and thermal dust emissions in the frequency
  range from 33 to 353 GHz, where most of the CMB studies are performed
  and where we expect a mixture of these two main foreground emissions. This model can
be very helpful to
  estimate the contamination by foregrounds of the polarized CMB
  anisotropies, for experiments like the \Planck\ satellite.
\end{abstract}


\end{frontmatter}



\section{Introduction}
\label{introduction}

\indent The \Planck\footnote{http://www.rssd.esa.int/index.php?project=Planck} satellite mission (\cite{early1, tauber}), currently in flight, will
provide measurements of the CMB anisotropies both in
temperature and polarization over the full sky with an unprecedented
accuracy~\citep{bluebook}. In particular, it should be able to measure
the tensor-to-scalar ratio, $r$, which sets the energy scale of inflation (\cite{lyth,peiris}).
For a extended mission and in the case of no direct detection, 
Planck should set an upper limit of $r < 0.03$ \citep{burigana, efstathiou1,efstathiou2}),and thus provides 
tight constraints on inflationary models (\cite{baumann}). To achieve this
high level of sensitivity, it is necessary to
estimate accurately the temperature and polarization anisotropies
from foreground diffuse Galactic emissions
and from point-like and compact sources of Galactic and extraGalactic
origin. A reliable estimation of the residual contamination due
to the foreground emissions after application of component separation
methods (see \cite{betoules2009, leach} for recent
studies) is thus necessary to retrieve the
cosmological information from the \Planck\ data.\\

\indent As summarized in~\cite{fauvet} the main polarized foreground
contributions come from the diffuse Galactic synchrotron emission \citep{page2007}
and from thermal dust emission \citep{benoit2004a, ponthieu2005}. The polarized
synchrotron emission have already been modelled by \cite{page2007,sun,jaffe,jansson,fauvet} based on
models of the Galactic magnetic field~\citep{han2004,han2006} and of
the relativistic electrons in the Galaxy. Concerning the thermal dust
emission, modeling based on the physical origins of this polarized emission
has already been discussed in~\cite{ponthieu2005,page2007,fauvet}.\\

\indent We propose here an extended analysis of the 3D joint model of
the Galactic polarized diffuse
emissions discussed in~\cite{fauvet}. In the previous analysis we
focused on the \Wmap\ data at 23 GHz and \Archeops\ at 353 GHz data
where synchrotron and thermal dust emissions dominate, respectively.
Here, we use complementary data : the other frequency bands of \Wmap\ from 33
to 94 GHz, where a mixing of those
emissions is expected, and the \Archeops\ data at 353 GHz. Furthermore
we apply here a
pixel-to-pixel likelihood based comparison instead of a Galactic
profile-based method as discussed in~\cite{fauvet}.\\

\indent The paper is structured as follows: Sect.~\ref{data} describes the
five-year \Wmap\ and \Archeops\ data set used in the analysis. In
Sect.\ref{buil_map} we describe in detail models for the polarized
foreground emissions. Section~\ref{3dgal_model} discusses the 3D model
of the Galaxy using to construct the polarized Galactic emissions.
The models are statistically compared to the data in Section~\ref{pix_comp}
and we discuss the results in Section~\ref{result}. We finally conclude in Section~\ref{conc}.

\section{Observations}
\label{data}

\subsection{Diffuse Galactic synchrotron emission}

\indent The synchrotron emission is an important contributor to the
diffuse sky emission at both radio and microwave observation
frequencies.  \\

\indent In intensity, the 408MHz all-sky continuum survey (\cite{haslam}), at a resolution of  of 0.85 degrees, is a good tracer of the  synchrotron emission and it will be used in the following
as a template. In particular, we use the 408MHz all-sky map available on the LAMBDA website in the
\HEALpix\ pixelisation scheme~\citep{gorski}. We correct this map from the contribution from free-free emission which estimated to be about 30~\%. The corrected maps was then
downgraded to the resolution of the WMAP and Archeops maps discussed below. \\

\indent In polarization, Faraday rotation introduces complications into the interpretation of the radio data
since strong depolarization is observed for frequencies lower than
10~GHz, particularly concerning the inner part of the Galaxy
(\cite{burn1966,sun,jaffe,jansson,2006A&A...455L...9L}). A detailed discussion on the depolarization process, based
on data from the Leiden survey, can be found in~\cite{porta2006}. For this reason the best polarized Galactic diffuse synchrotron tracers
are at high frequency such as the \Wmap\ survey (\cite{page2007}).\\

\indent We used here the five-year \Wmap\ Q and U low resolution maps
for the frequency channels centered at
33 (Ka band), 41 (Q band), 61 (V band) and 94~GHz (W band)
(\cite{page2003,gold}). These data are available
on the LAMBDA website in the
\HEALpix\ pixelisation scheme at $N_{side}=16$.
The associated noise is estimated using the full
noise correlation matrix also available on the LAMBDA website in the
same resolution.

\subsection{Thermal dust}

\indent The thermal dust emission is significant in the \Wmap\ data only for
frequencies above 70 GHz, then we also used here the \Archeops\
353 GHz Q and U maps as tracers of the polarized thermal dust
emission. Those maps cover about 20 \% of the sky~\citep{macias} and were filtered and downgraded to $N_{side} = 16$ to make them
comparable to the \Wmap\ ones. \\

In intensity, the most accurate measurements of the thermal dust emission are those of the IRAS satellite~\citep{neugebauer} and in particular 
at 100~$\mu$m. We use here predicted full-sky maps of sub millimeter and microwave emission from the diffuse interstellar dust in the Galaxy
from~\cite{finkbeiner} which were produced combining the IRAS data at 6.1 arcmin and the COBE DIRBE data at 40 arcmin~\citep{schlegel}.
These maps were downgraded to the resolution of the WMAP and Archeops maps presented above.

\section{Emissivity model in polarization}
\label{buil_map}

\indent We present in this section a realistic model of the diffuse
polarized synchrotron and dust emissions
using a 3D model of the Galactic magnetic field and of the matter
density in the Galaxy. We will
consider the distribution of relativistic cosmic-ray electrons (CREs),
$n_\mathrm{CRE}$, for the synchrotron emission and the distribution of
dust grains, $n_{dust}$, for the thermal dust
emission. Following~\cite{fauvet} we calculate the Stokes parameters
I, Q and U for the Galactic polarized emission along the line of sight
as follows.\\

\indent For the synchrotron emission (\cite{ribicki}) we write :
{\small
\begin{eqnarray}
\label{eq:map_sync}
\centering
 I^{\mathrm{sync}}_{\nu}(\mathrm{{\bf n}}) &=&
I^{\mathrm{Has/ff}}(\mathrm{{\bf n}})
\left(\frac{\nu}{0.408}\right)^{\beta_s},\\
Q^{\mathrm{sync}}_{\nu}(\mathrm{{\bf n}}) &=& I_{\mathrm{Has/ff}}(\mathrm{{\bf
n}})
\left(\frac{\nu}{0.408}\right)^{\beta_s} \\
&& \frac{\int \cos(2\gamma({\bf n},s))p_s
  \left(B_l^2({\bf n},s) + B_t^2({\bf n},s)
\right)\mathrm{n}_{\mathrm{CRE}}(\mathrm{{\bf n}},s)ds}{\int \left(B_l^2({\bf n},z) +
B_t^2({\bf n},s) \right)\mathrm{n}_{\mathrm{CRE}}(\mathrm{{\bf n}},s)ds},\\
U^{\mathrm{sync}}_{\nu}(\mathrm{{\bf n}}) &=& I_{\mathrm{Has/ff}}(\mathrm{{\bf
n}})
\left(\frac{\nu}{0.408}\right)^{\beta_s} \\
&&\frac{\int \sin(2\gamma({\bf n},s))p_s
  \left(B_l^2({\bf n},s) + B_t^2({\bf n},s)
\right)\mathrm{n}_{\mathrm{CRE}}(\mathrm{{\bf n}},s)ds}{\int \left(B_l^2({\bf n},s) +
B_t^2({\bf n},s) \right)\mathrm{n}_{\mathrm{CRE}}(\mathrm{{\bf n}},s)ds},
\end{eqnarray}
}
\noindent where $B_n(\mathrm{{\bf n}},s)$ is the magnetic component
along the line-of-sight {\bf n}, and $B_l(\mathrm{{\bf n}},s)$ and
$B_t(\mathrm{{\bf n}},s)$ the magnetic field components
on a plane perpendicular to the line-of-sight. 
Notice that the 3 vectors {\it n,l,t} form an orthonormal basis being 
{\it l} and {\it t} oriented to the north and to the east respectively in a plane perpendicular to {\it n}. The polarization fraction
$p_s$ is set to 75\%~\citep{ribicki}. The polarization angle $\gamma({\bf n},s)$ is
given by :

\begin{eqnarray}
\centering
 \gamma(\mathrm{{\bf n}},s) &=& \frac{1}{2}
\arctan{\left(\frac{2B_l(\mathrm{{\bf n}},s) \cdot B_t(\mathrm{{\bf
n}},s)}{B^2_l(\mathrm{{\bf n}},s) -B^2_t(\mathrm{{\bf n}},s)} \right)}.
\end{eqnarray}

\indent The distribution of relativistic electrons,
$\mathrm{n}_{\mathrm{CRE}}$, is described in detail in
section~\ref{3dgal_model}. $I_{\mathrm{Has/ff}}$ is the reference map in
intensity
constructed from the \emph{408 MHz all sky continuum
  survey}~\cite{haslam} after subtraction of the bremsstrahlung
(\emph{free-free}) emission
and $\nu$ is the frequency of observation. To subtract the free-free
contribution
we used the \Wmap\ K-band free-free foreground map generated from the
maximum entropy method (MEM)~\citep{hinshaw, bennett2003a}. Notice that
we do not use the synchrotron MEM intensity map at 23~GHz
(\cite{hinshaw}) as a synchrotron template to avoid any possible
Anomalous Microwave Emission (AME) contamination (the \Wmap\ team made no attempt to fit for
the latter). The spectral index $\beta_s$ used to extrapolate maps at various
frequencies is a free parameter of the model.  {\bf The SED of the synchrotron emission in the radio and microwave domain,
and in particular in the 33 to 353~GHz range, can be well approximated by a power law in antenna temperature units (\cite{ribicki}).
This is due to the fact that the energy spectrum of the Galactic relativistic electrons producing the radio and microwave synchrotron emission is also well approximated by a power law
(\cite{2004ApJ...601L..13K}). }. \\

\indent For the thermal dust emission we write
\begin{eqnarray}
\centering
 I^{\mathrm{dust}}_{\nu}(\mathrm{{\bf n}}) &=&
\mathrm{I}_{\mathrm{fds}}(\mathrm{{\bf n}}) \left(\frac{\nu}{353}
\right)^{\beta_d},\\
Q^ {\mathrm{dust}}_{\nu}(\mathrm{{\bf n}}) &=&
\mathrm{I}_{\mathrm{fds}}(\mathrm{{\bf n}}) \left(
  \frac{\nu}{353}\right)^{\beta_d}\\
&& \frac{\int \cos(2 \gamma(\mathrm{{\bf n}},s))
  \sin^2(\alpha) f_{\mathrm{norm}}p_d
\mathrm{n}_{\mathrm{dust}}(\mathrm{{\bf n}},s)ds}{\int
\mathrm{n}_{\mathrm{dust}}(\mathrm{{\bf n}},s)ds},\\
U^{\mathrm{dust}}_{\nu}(\mathrm{{\bf n}}) &=&
\mathrm{I}_{\mathrm{fds}}(\mathrm{{\bf n}})
\left(\frac{\nu}{353}\right)^{\beta_d} \\
&& \frac{\int \sin(2 \gamma(\mathrm{{\bf n}},s))
  \sin^2(\alpha) f_{\mathrm{norm}}p_d
\mathrm{n}_{\mathrm{dust}}(\mathrm{{\bf n}},s)ds}{ \int
\mathrm{n}_{\mathrm{dust}}(\mathrm{{\bf n}},s)ds} ,
\end{eqnarray}

\noindent where the dust polarization fraction $p_d$ is set to 10
\%~\citep{ponthieu2005} based on the \Archeops\ data, and $\mathrm{n}_{\mathrm{dust}}(r,z)$ is the dust
grain distribution discussed in section~\ref{3dgal_model}. The
$sin2(\alpha)$ term accounts for
the geometrical suppression and $f_{norm}$ is an
empirical factor which accounts for the misalignment between dust
grains and the magnetic field lines (see~\citep{fauvet} for details).
The reference map, $\mathrm{I}_{\mathrm{fds}}$ was taken to be model 8 in \cite{finkbeiner} at 545~GHz. The spectral index $\beta_d$ used to
extrapolate maps at various frequencies is a free parameter of the model. 
In the following we work on antenna temperature, Rayleigh-Jeans units. Assuming nearly constant dust temperature 
across the sky then a power-law approximation for the thermal dust emission in antenna temperature
units can be used~\citep{planck}.\\

\section{A 3D modeling of the Galaxy}
\label{3dgal_model}

\indent We describe here the 3D model of the Galaxy as used in the
previous Stokes parameter definitions both for synchrotron and dust.

\subsection{Matter density model}

\indent In galactocentric cylindrical coordinates $(r,z,\phi)$ we
consider an exponential distribution of relativistic electrons
$n_{CRE}$ on the Galactic disk motivated by \cite{drimmel}:

\begin{eqnarray}
\centering
 n_\mathrm{CRE}(r,z) &=& n_{0,e} \cdot
\frac{e^{-\frac{r}{n_{\mathrm{CRE},r}}}}{\cosh2(z/n_{\mathrm{CRE},h})},
\end{eqnarray}

\noindent where $n_{\mathrm{CRE},h}$ defines the width of the distribution
vertically and it is set to $1$ kpc in the
following. $n_{\mathrm{CRE},r}$ defines the distribution radially and
it is set to $3$ kpc (see \cite{sun,jaffe,fauvet}). The value of
$n_{0,e}$ is set to $6.4\times 10^{-6}$.cm$^{-3}$~\citep{sun}. \\

\noindent The density distribution of dust grains in the Galaxy is
poorly known and we therefore choose to describe it in the same way as
for relativistic electrons:

\be
\centering n_d(r,z) = n_{0,d} \cdot
\frac{e^{-\frac{r}{n_{d,r}}}}{\cosh2(z/n_{d,h})},
\ee

\noindent where $n_{d,r}$ and $n_{d,h}$ are the radial and vertical
widths of the distribution.
In the following we set them to 3 and 1 kpc respectively. Notice that we expect these two
parameters to be strongly correlated for both dust grain and
  electron distributions, hence we decided to fix one of
them as in previous analyses (\cite{sun,jaffe}). We have tested
different
values of these two parameters and found no impact on the final results.

\subsection{Galactic magnetic field model}
\label{mg_field_model}

\indent  The Galactic magnetic field model consists of a regular component and a turbulent component such that
\be
\centering
{\bf B}_{tot} (\mathbf {r}) = {\bf B}_{reg}(\mathbf {r}) + A_{turb} \  {\bf B}_{turb}(\mathbf {r})
\ee
{\bf where $A_{turb}$ is the amplitude of the turbulent component and it is a free parameter of the model. 
In the following we will express it as a relative amplitude with respect to the amplitude of the regular component.}

\subsubsection{Regular component}

\noindent The regular part ${\bf B}_{reg}(\mathbf {r})$ is a Modified
Logarithmic Spiral model, discussed in detail in~\cite{fauvet}. In
galactocentric cylindrical coordinates
$(r, \phi, z)$ it reads

\begin{eqnarray}
\centering
\mathbf{B}(\mathbf {r})&=& B_{reg}(\mathbf {r})[ \cos(\phi+\beta) \ln
\left( \frac{r}{r_0} \right) \sin(p) \cos(\chi ) \cdot \mathbf{u_r}
\nonumber \\&&- \cos(\phi+\beta) \ln \left( \frac{r}{r_0} \right)
\cos(p) \cos(\chi) \cdot \mathbf{u_{\phi}}  \nonumber \\&&+ sin(\chi)
\cdot \mathbf{u_z}] ,
\end{eqnarray}

\noindent where $p$ is the pitch angle, $\beta=1/\tan(p)$ and $r_0$
is the radial scale set to 7.1 kpc. $\chi(r) = \chi_0(r)(z/z_0)$ is the
vertical
scale, with $\chi_0 = 22.4$ degrees and $z_0 = 1$ kpc.
Following \cite{taylor} we restrict our model to the range $3 < r < 20$
kpc. The lower limit is set to avoid the center of the Galaxy
for which the physics is poorly constrained and the model diverges. The
intensity of the regular field is fixed using pulsar Faraday rotation
measurements by \cite{han2006}
\be
B_{reg}(r) = B_0 \ e^{-\frac{r-R_{\odot}}{R_B}}
\ee

\noindent where the large-scale field intensity at the Sun position is
$B_0=2.1 \pm 0.3 \mu
G$ and the physical cut $R_B = 8.5 \pm 4.7$ kpc. The distance
between the Sun and the Galactic center, $R_{\odot}$ is set to 8 kpc
(\cite{eisenhauer, reid}). \\

\subsubsection{Turbulent component}
\label{turb_comp}

\indent In addition to the large-scale Galactic magnetic field,
Faraday rotation measurements on pulsars in our vicinity have revealed
a turbulent component on scales smaller than a few hundred pc
(\cite{lyne}). Moreover it seems to be present on large angular scales
(\cite{han2004}) with an amplitude estimated to be of the same order of magnitude as that of the regular one (\cite{han2006}). 
The magnetic energy in Fourier space, $E_B(k)$, associated with the turbulent component is well described by a power spectrum of the form (\cite{han2004, han2006})
\be       
\centering
E_B(k) = C \left(\frac{k}{k_0}\right)^{\alpha}
\label{eq_pw_ko}
\ee

\noindent where $\alpha = -0.37$ and $C = (6.8 \pm 0.3) \cdot 10^{-13}\,\mathrm{erg\,cm^{-3}\,kpc}$. 
To obtain the 3D spatial distribution of the turbulent magnetic field we produced independent Gaussian simulations
from the above power spectrum in the $x$, $y$ and $z$ directions on boxes of of 512$^3$ pixels at a resolution of 56 pc.
We renormalize the simulated boxes so that the total amplitude of the turbulent component is $A_{turb}$. \\

\noindent Notice that we do not include here an anisotropic/ordered
component as discussed in~\cite{jaffe}. As in \citep{fauvet}, our regular
component is then equivalent to the sum of what \cite{jaffe} call the
coherent and ordered fields. The latter, also called ordered random
component, will be not considered in this paper because it can not be
distinguished from the large-scale magnetic field when studying
polarization intensity only. \\


\section{Method}
\label{pix_comp}

\indent We compute I, Q and U maps for the synchrotron and thermal dust emissions with
a modified version of
the \Hammurabi\ code (\cite{waelkens}). Each map is generated by
integrating in 100 steps along each line-of-sight defined
by the \HEALpix \citep{gorski} $N_\mathrm{side}=16$ pixel centres. The integration
continues out to 25~kpc from the observer situated
8.5~kpc from the Galactic centre. These full-sky maps are computed for a
grid of models obtained by varying the pitch angle, $p$,
the turbulent component amplitude, $A_{turb}$ and the spectral indices
of the synchrotron and thermal dust emissions $\beta_s$ and $\beta_d$.
The latter are assumed to be spatially constant on the sky. Dealing with
a more realistic
varying spectral index (see \cite{kogut,porta2008} for
detailed studies) is beyond the scope of this paper. However, we ensured
that this hypothesis does not impact the results for the other free
parameters in the model. Indeed, we produced
simulated \Wmap\ observations with spatially varying synchrotron
spectral index and analyzed them assuming a constant one.
No significant bias was observed for any of the other parameters and the
error bars were equivalent to those in the case of a constant spectral index. \\

\indent The range and binning step considered for each of the above
parameters are given in Table~\ref{tab:param_pixpix_tab}.
All the other parameters of the models of the Galactic magnetic field
and matter density are fixed to values proposed
in Section~\ref{3dgal_model}. Notice that to be able to compare the dust
models to the \Archeops\ 353~GHz data, the simulated maps
are multiplied by a mask to account for the \Archeops\ partial sky
coverage of 30\% \citep{macias}. \\

\begin{table}
\begin{center}
\caption{\footnotesize Parameters of the 3D Galactic diffuse emissions
  model.\label{tab:param_pixpix_tab}}
\vspace{0.3cm}

\begin{tabular}{|c|c|c|} \hline
 Parameter  &   Range   &   Binning  \\\hline
$p$ (deg) &    $[-80.0,15.0]$  &   $5.0$ \\\hline
$\beta_s$ &    $[-4.5, -2.8]$  & $0.05$  \\\hline
$\beta_d$ &    $[0.05, 2.5]$  & $0.05$  \\\hline
$A_{turb}$&    $[0.1, 1.1]*B_{reg}$  & $0.1$  \\\hline
\end{tabular}
\end{center}
\end{table}

\subsection{Likelihood based analysis}

\indent To compare the \Wmap\ data sets and the model of Galactic polarized
emissions we used a maximum likelihood analysis where
the the log-likelihood function is given by

\be
- \log \mathcal{L}_{\nu} = \sum_{i}
\sum_{j=0}^{N_{\mathrm{pix}}-1}(D^{\nu}_{i,j}-M^{\nu}_{i,j})N^{-1}_{\mathrm{inv},\nu}(D^{\nu}_{i,j}-M^{\nu}_{i,j})
\ee

\noindent where $D_{i,j}^{\nu}$ and $M_{i,j}^{\nu}$
correspond respectively to the data set and model for the frequency of observation
$\nu$. $i$ and $j$ index the polarization states Q and U and the pixel
number in the maps respectively. $N^{-1}_{\mathrm{inv},\nu}$ is the
pixel-to-pixel inverse covariance matrices for the frequency band
$\nu$. These matrices allow us to estimate the noise for a given pixel and
the correlation between pixels. They are composed of
6144$\times$6144 elements allocated in 4 blocks of 3072$\times$3072
elements each. Each element represents the auto-correlated noise
associated to a
pixel or the correlation between 2 pixels. The elements linked to the
pixel located outer the processing mask are set to
0~\citep{hinshaw09,limon}. These matrices are available on the LAMBDA website.\footnote{http://lambda.gsfc.nasa.gov/product/map/current/} \\

\indent We used the processing mask of the \Wmap\
team. This processing mask is built from intensity cuts  on the \Wmap\
polarized maps at 23 GHz and the thermal dust emission model of the
\Wmap\ team~\citep{kogut, page2007}. The point-sources have also been
subtracted. More details concerning this mask can be found in~\cite{hinshaw09,limon}.\\

\indent We also used a maximum likelihood analysis to compare the
thermal dust
emission model to the \Archeops\ data. The log-likelihood function is
defined by

\be
- \log \mathcal{L} = \sum_i \sum^{N_\mathrm{pix}-1}_{j=0}
\frac{(D_{i,j}-M_{i,j})(D_{i,j}-M_{i,j})}{\sigma2_{i,j}}
\ee

\noindent where $D_{i,j}^{\nu}$ and $M_{i,j}^{\nu}$
correspond to the \Archeops\ data and to the thermal dust emission
model at 353 GHz. $i$ and $j$ index the Stokes parameters Q and U and
the pixel number respectively. The error bars associated to the
\Archeops\ data are estimated from 600 simulations of the time ordered
data (TOD) (see~\cite{ponthieu2005, macias}).  The instrumental noise at the TOD level was estimated
following~\cite{benoit2003a}. From these simulations we conclude that the noise 
in the Archeops maps can be well approximated by anisotropic white noise on the maps. 
We thus compute the variance per pixel.

\section{Results}
\label{result}

\begin{table*}
\begin{center}
\caption{\footnotesize Best-fit parameters for the models of the Galactic
  polarized emissions including a MLS Galactic magnetic field,
  constrained using the \Wmap\ and \Archeops\ data inner and outer the
P06 mask. The $\chi^2$ are given by degree of freedom} \label{tab:param_wmap}
\vspace{0.3cm}
\hspace*{-1cm}
\begin{tabular}{|c|c|c|c|c|c|} \hline
Frequency band & zone      & $p (deg)$  & $\beta_s$ & $\beta_d$  &
$\chi^2_{min}$  \\\hline
$Ka$               &   in P06  & $ -30.0^{+13}_{-20}$   &
$-3.45 \pm 0.5$ & $0.5^{+0.3}_{-0.1}$ & $10.09$   \\\cline{2-6}
                    &  out of P06 & $ -35.0^{+10}_{-15}$   &
$-3.5^{+0.05}_{-0.65}$ &$0.5^{+0.9}_{-0.15}$  & $2.89$ \\\cline{2-6}
                     & all-sky & $ -35.0^{+20}_{-25}$   &
$-3.45^{+0.05}_{-0.8}$ &$0.45^{+0.9}_{-0.25}$  & $10.08$ \\\hline
$ Q $                & in P06 & $-30.0^{+25}_{-20}$ &
$-3.45^{+0.05}_{-0.8}$ &$0.8^{+0.8}_{-0.2}$ & $3.65$ \\\cline{2-6}
                     & out of P06 & $-30.0^{+20}_{-15}$ &
$-3.65^{+0.25}_{-0.05}$ & $0.8^{+0.7}_{-0.4}$ &$1.58$  \\\cline{2-6}
                     & all-sky & $-30.0^{+20}_{-15}$ &
$-3.45^{+0.2}_{-0.7}$ & $0.8^{+1.3}_{-0.2}$ &$3.67$  \\\hline
$ V $                & in P06 & $ -15.0^{+10}_{-17}$  &
$-3.4^{+0.15}_{-0.8}$ &$1.25^{+0.9}_{-0.4}$ & $1.26$ \\\cline{2-6}
                     & out of P06 & $-5.0^{+5}_{-40}$   &
$-3.95^{+0.7}_{-0.4}$ &$1.8^{+0.3}_{-0.9}$ & $1.07$ \\\cline{2-6}
                     & all-sky & $-15.0^{+10}_{-25}$   &
$-3.4^{+0.15}_{-0.85}$ &$1.25^{+0.7}_{-0.8}$ & $1.26$ \\\hline
$ W $                & in P06 & $-60.0^{+35}_{-15}$  &
$-3.2^{+0.15}_{-1.0}$ & $ 1.56^{+1.05}_{-0.15}$ & $1.32$ \\\cline{2-6}
                     & out of P06 & $-35.0^{+30}_{-25}$   &
$-3.7^{+0.2}_{-0.6}$ &$2.15^{+0.25}_{-0.05}$ & $1.2$  \\\cline{2-6}
                     & all-sky & $-60.0^{+30}_{-15}$   &
$-3.2^{+0.2}_{-1.0}$ &$1.5^{+0.7}_{-0.6}$ & $1.32$  \\\hline
 all \Wmap\ bands    & all-sky & $-30.0^{+25}_{-10}$   &
 $-3.45^{+0.1}_{-0.4}$ &$1.0^{+0.9}_{-0.2}$ &  $16.30$ \\\hline
\Archeops\ 353 GHz & 30 \% sky & $-35.0^{+15}_{-10}$ & $\emptyset$ &
$1.8^{+0.4}_{-0.3}$ & $1.105$  \\\hline

\end{tabular}
\end{center}
\end{table*}

\begin{figure*}
\centering
\includegraphics[height=10cm,width=10cm]{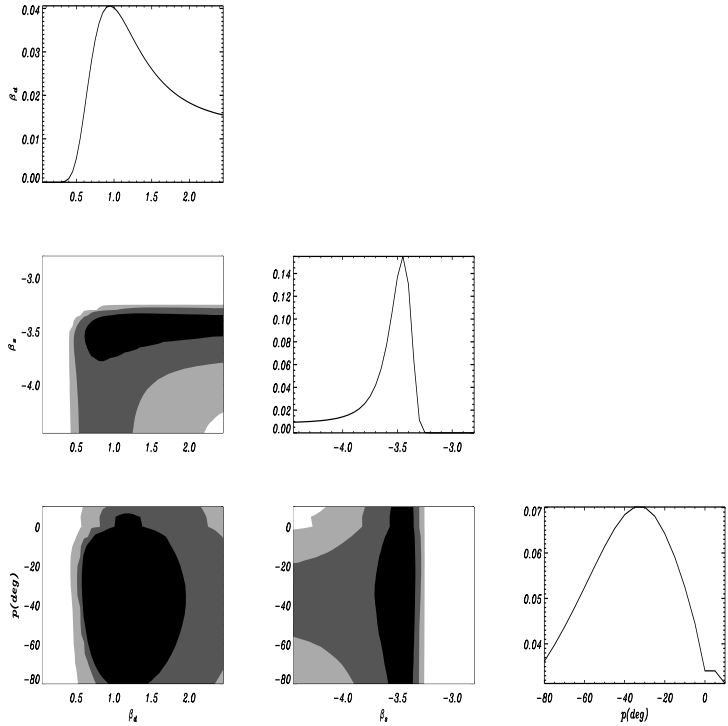}\caption{\footnotesize
 Combined 1 and 2D marginalized likelihood, using all the \Wmap\ channels, for the parameters $A_{turb}$, $\beta_s$,
$\beta_d$ and $p$ assuming no turbulent magnetic field component. We present
  the 68.8\% (dark), 95.4\% (dark grey) and 98\% (grey) confidence level contours. \label{fig:likeKa_fturb}}
\end{figure*}

\begin{figure*}
\centering
\includegraphics[height=10cm,width=10cm]{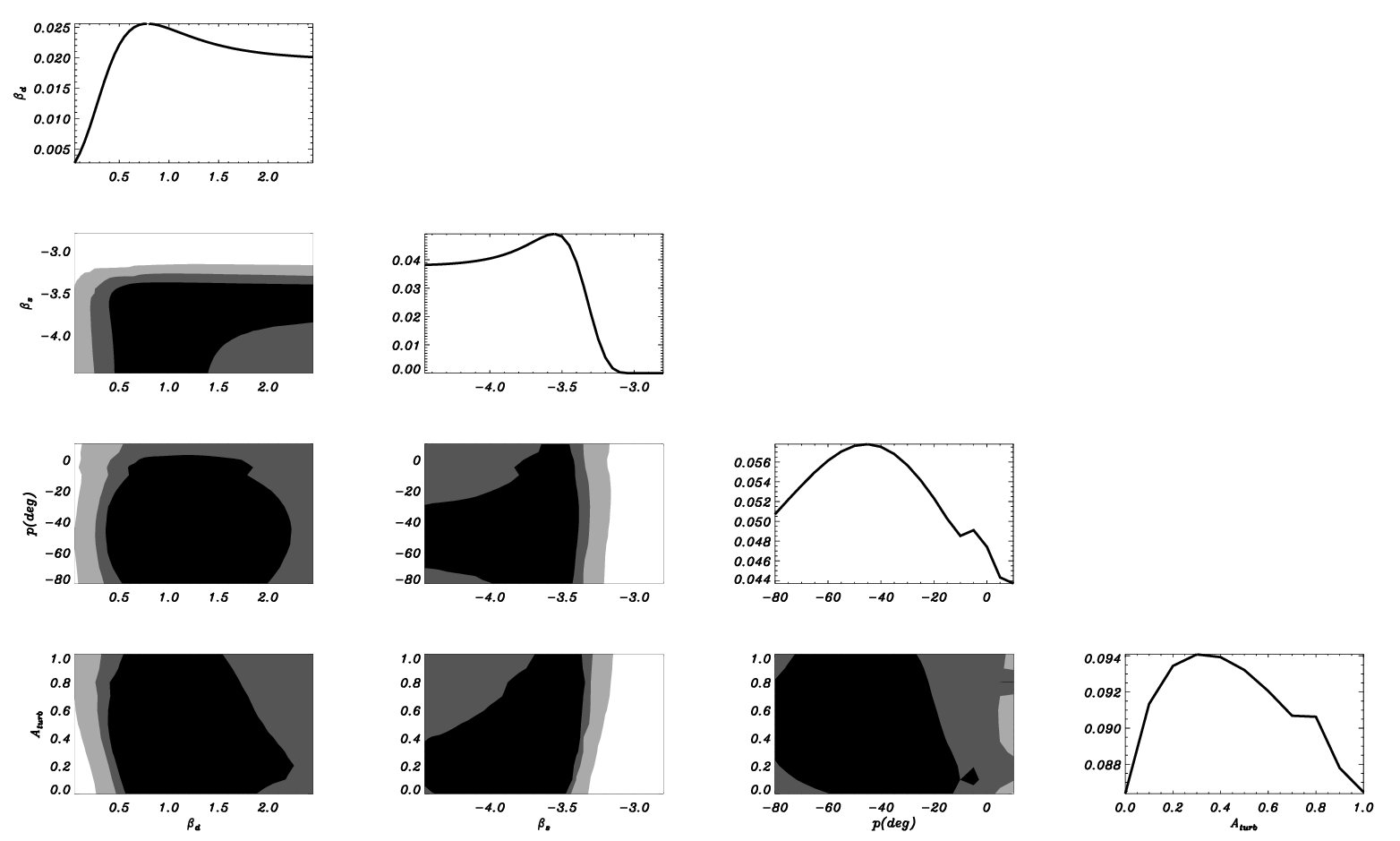}\caption{\footnotesize
  Combined 1 and 2D marginalized likelihood, using all the \Wmap\ channels, for the parameters $A_{turb}$, $\beta_s$,
$\beta_d$ and $p$ including a turbulent magnetic field. We present the 68.8\% (black),  95.4\% (dark grey) and 98\% (grey) confidence level contours. 
  \label{fig:like_fturb}}
\end{figure*}

\indent First of all, we consider a null turbulent magnetic field component.
For this case, the constraints on the parameters of the polarized Galactic
emission model, $\beta_s$, $\beta_d$ and $p$, using the \Wmap\ and Archeops data are presented in Table~\ref{tab:param_wmap}. 
For illustration, we also present on Figure~\ref{fig:likeKa_fturb} the combined marginalized likelihood 
at 1 and 2D using all the \Wmap\ data. We present the 68.8\% (black), 95.4\% (dark grey) and 98\% (grey) confidence level contours.
The pitch angle, $p$, of the regular magnetic field component does not seem to be correlated neither with $\beta_s$ nor $\beta_d$.
Similar results are found in the case of the Archeops data. The best-fit values obtained for magnetic field pitch angle, $p$, are compatible for the \Wmap\ and \Archeops\ data,
indicating that the same magnetic field can describe both emissions. For $\beta_d$, as it might be expected, constraints are only reliable at high frequency (94 and
353~GHz) and are compatible within the 1-$\sigma$ error bars. These results are also compatible at
1$\sigma$ level with results presented in~\cite{boulanger,gold}. Notice that the value obtained is for a fix degree of polarization. 
An independent determination of the degree of polarization and of the dust spectral index is not possible given
the available data. Indeed, both parameters act as multiplicative factors.  In order to discriminate between them we would
need a larger frequency coverage and sampling.The $\beta_s$ parameter is well constrained at the \Wmap\ frequencies, with a best-fit value of $\beta_s =-3.4^{+0.2}_{-0.5}$, using all
the frequency bands and it is consistent across the frequency range. This result is  consistent at the 1$\sigma$ with the results obtained by~\cite{fauvet,gold, gold2011,page2007, sun,jansson}. \\

\noindent For the \Wmap\ data we computed the likelihood for the full sky, inside and outside the \Wmap\
processing mask. In the case of $\beta_{s}$, the results obtained inner and outer the mask are consistent with those from the full sky analysis within the error bars. 
Thus, we can conclude that the spectral variations are smaller than the error bars in the determination of $\beta_s$.
This seems to be consistent with the results from ~\cite{gold, gold2011} which found that the synchrotron spectral index is relatively
constant (with respect to our error bars) across the sky but for the Galactic plane where they found a significant increased. 
 Although the uncertainties on the dust spectral index are much larger our results seem to be consistent with those in ~\cite{early24} that favor a
 larger dust spectral index at high Galactic latitudes. \\

\begin{table*}
\begin{center}
\hspace*{-1cm}
\caption{\footnotesize Best-fit parameters for the models of the Galactic
  polarized emissions including a MLS Galactic magnetic field and a
  turbulent magnetic field.The $\chi^2$ are given by degree of freedom. \label{tab:param_rmodel_turb}}
\vspace{0.3cm}
\begin{tabular}{|c|c|c|c|c|c|} \hline
channel & $A_{turb}$ & $ p $ & $\beta_s $  & $\beta_d$ & $ \chi^2_{min}$
\\\hline
Ka & $0.3 \pm 0.2$  & $-45^{+27}_{-13}$   &  $-3.55^{+0.2}_{-0.5}$  &
$0.4^{+1.0}_{-0.2}$ & $13.081$ \\\hline
Q  & $0.8^{+0.1}_{-0.4}$  & $-40^{+25}_{-20}$ & $-3.55^{+0.2}_{-0.5}$ &
$0.7^{+0.8}_{-0.4}$ & $4.554$ \\\hline
V  & $<0.9$ (95.4 \% CL) & $0^{+5}_{-40}$ & $-3.6^{+0.3}_{-0.4}$ &
$1.25^{+0.6}_{-0.3}$ & $1.530$\\\hline
W  & $0.8^{+0.1}_{-0.4}$  & $-5^{+7}_{-37}$ & $-3.4^{+0.3}_{-0.5}$ &
$1.6^{+0.4}_{-0.8}$ & $1.463$\\\hline
all & $< 0.5$ (95.4 \% CL) & $-30^{+15}_{-17}$  & $-3.55^{+0.2}_{-0.5}$ &
$1.2^{+0.4}_{-0.3}$ & $20.0$\\\hline
\Archeops\ 353 GHz & $< 2.25$ (95.4 \% CL) & $-20^{+80}_{-50}$ & $\emptyset$ &
$1.8^{+0.7}_{-0.9}$ & $1.98$  \\\hline
\end{tabular}
\end{center}
\end{table*}

\noindent The constraints on the parameters of the polarized Galactic
emission model including a turbulent magnetic field component
are presented in Table~\ref{tab:param_rmodel_turb} for the \Wmap\ and \Archeops\ data. 
As above, the best-fit parameters of the model are consistent from 30 to 353 GHz but
the constraints are much more loosy. In particular, the relative amplitude of the turbulent component, $A_{turb}$, is poorly constrained
although the data do not seem to favour a strong turbulent component. We present on Figure~\ref{fig:like_fturb} the combined marginalized likelihood at 1 and 2D using all the \Wmap\ data. We present the 68.8\% (black), 95.4\% (dark grey) and 98\% (grey) confidence level contours.

\begin{figure*}
\begin{center}
\includegraphics[angle=90,height=4cm,width=7cm]{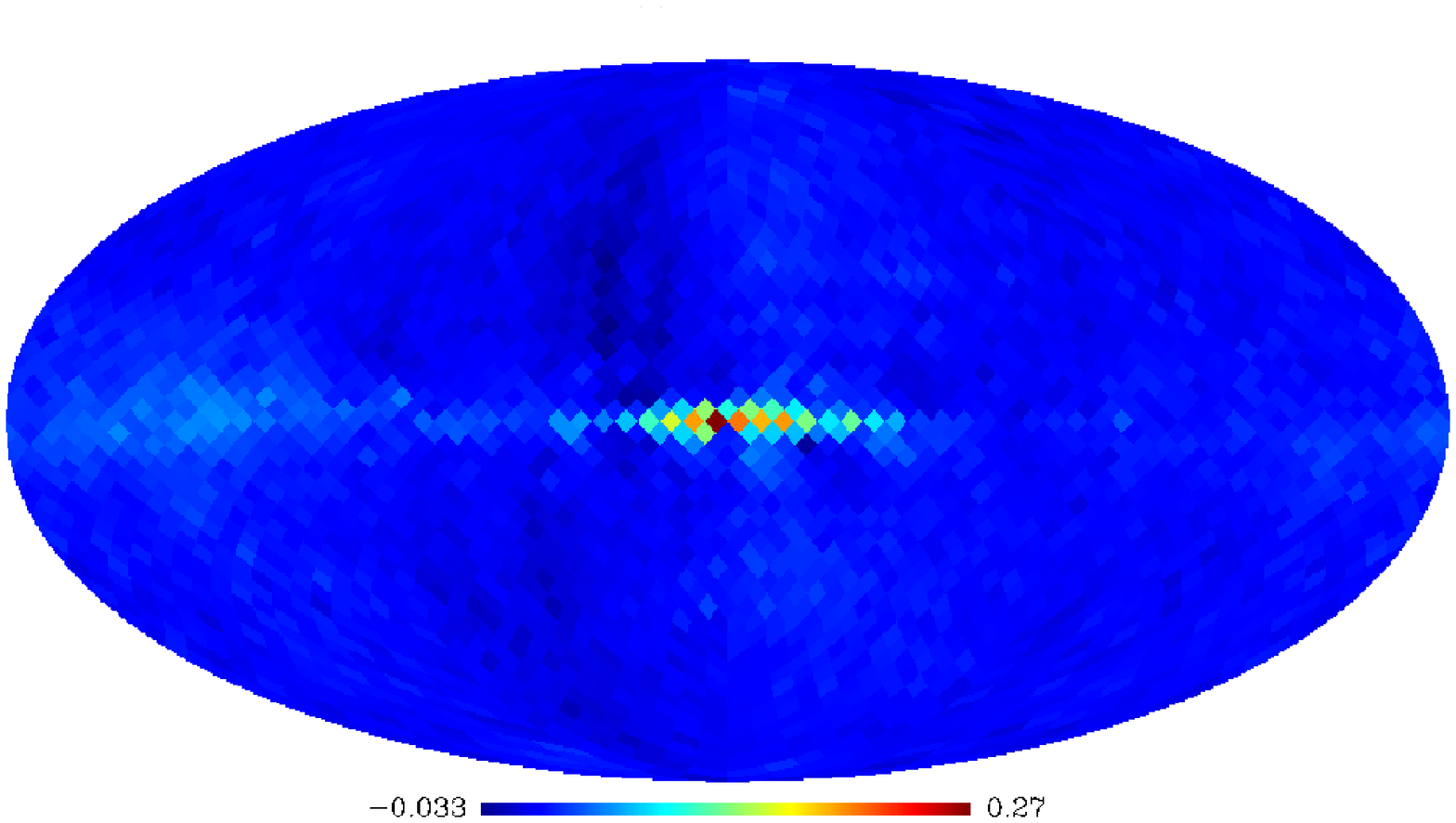}\includegraphics[angle=90,height=4cm,width=7cm]{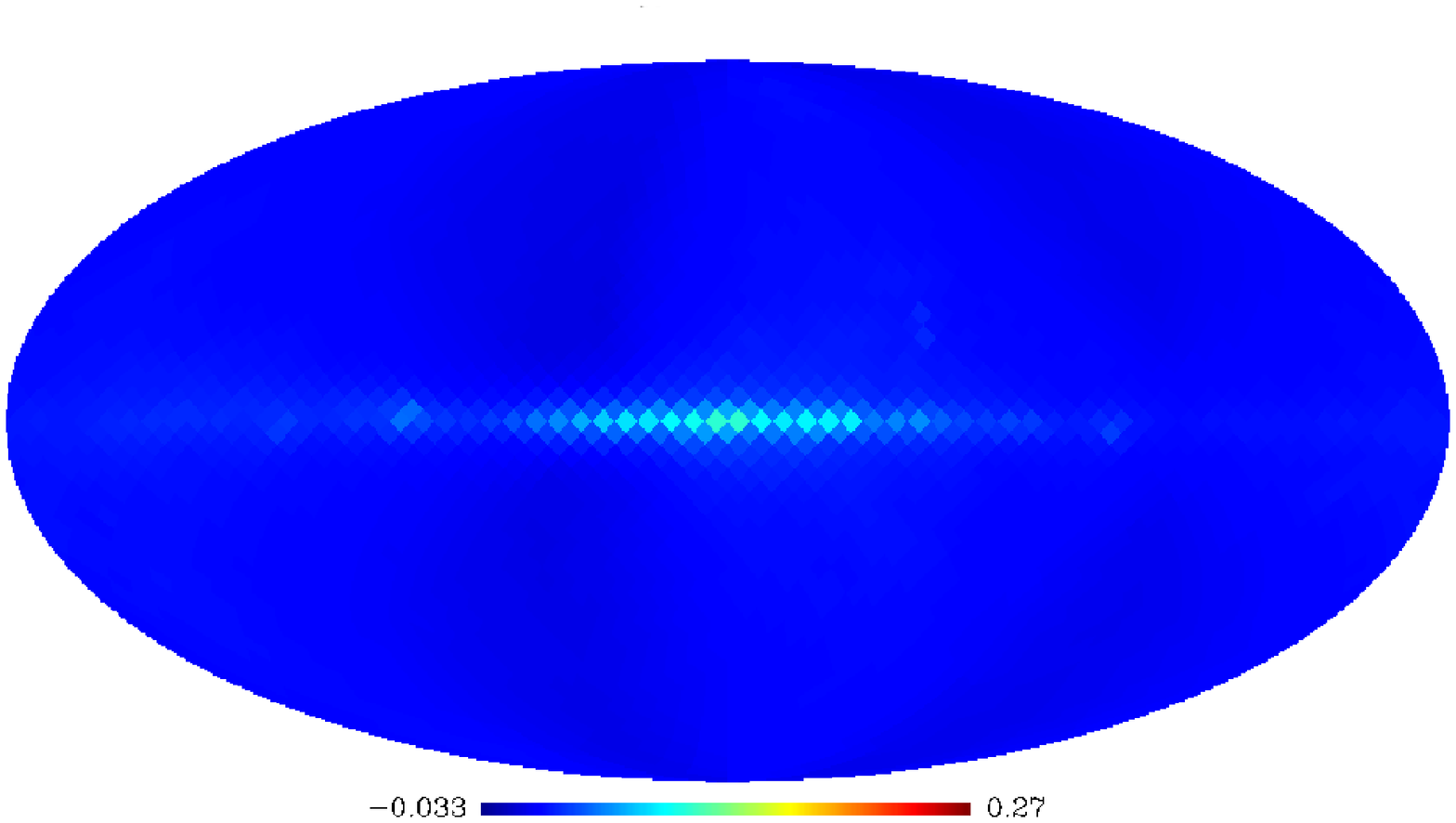}
\includegraphics[angle=90,height=4cm,width=7cm]{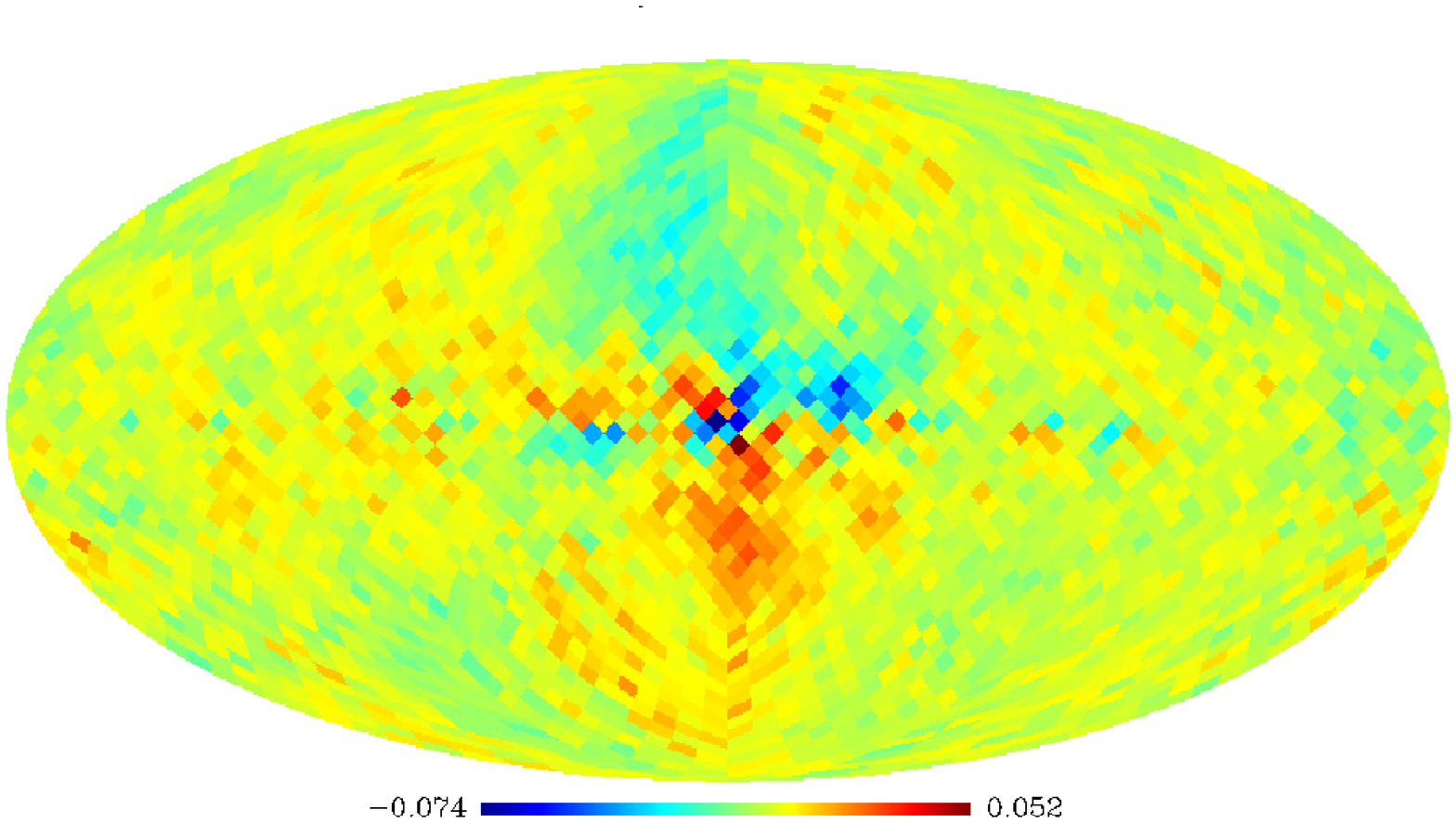}\includegraphics[angle=90,height=4cm,width=7cm]{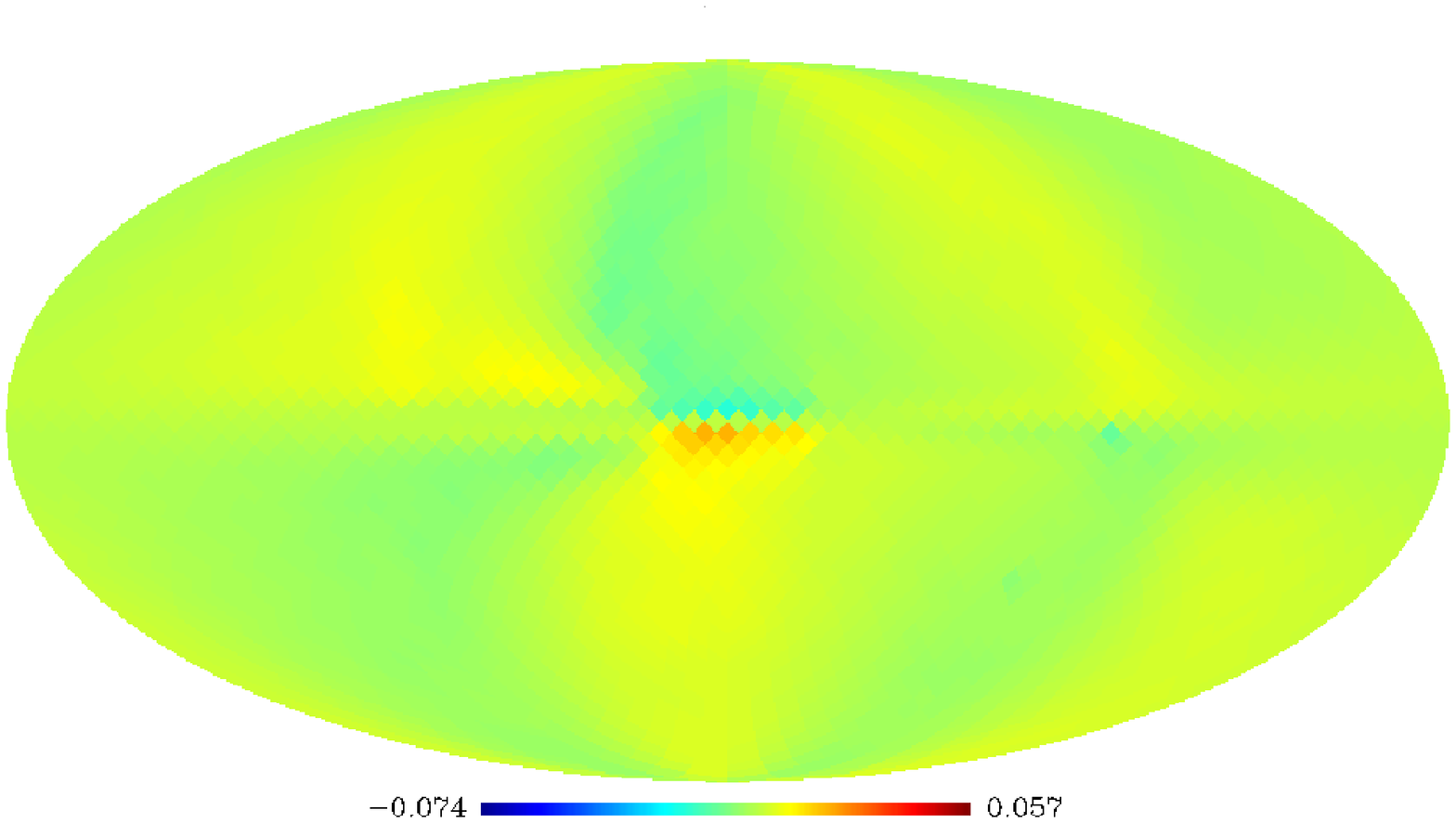}\caption{\footnotesize Maps (in $\mu K_{RJ}$) for the polarization Q (\emph{top}) and U(\emph{bottom}) Stokes parameters at 33 GHz for the \Wmap\ 5 years data (\emph{left}), and or the best-fit parameters model of the polarized foreground emissions (\emph{right}) .
\label{fig:mapQUpol_rmodel33}}
\end{center}
\end{figure*}

\begin{figure*}
\begin{center}
\includegraphics[angle=90,height=4cm,width=7cm]{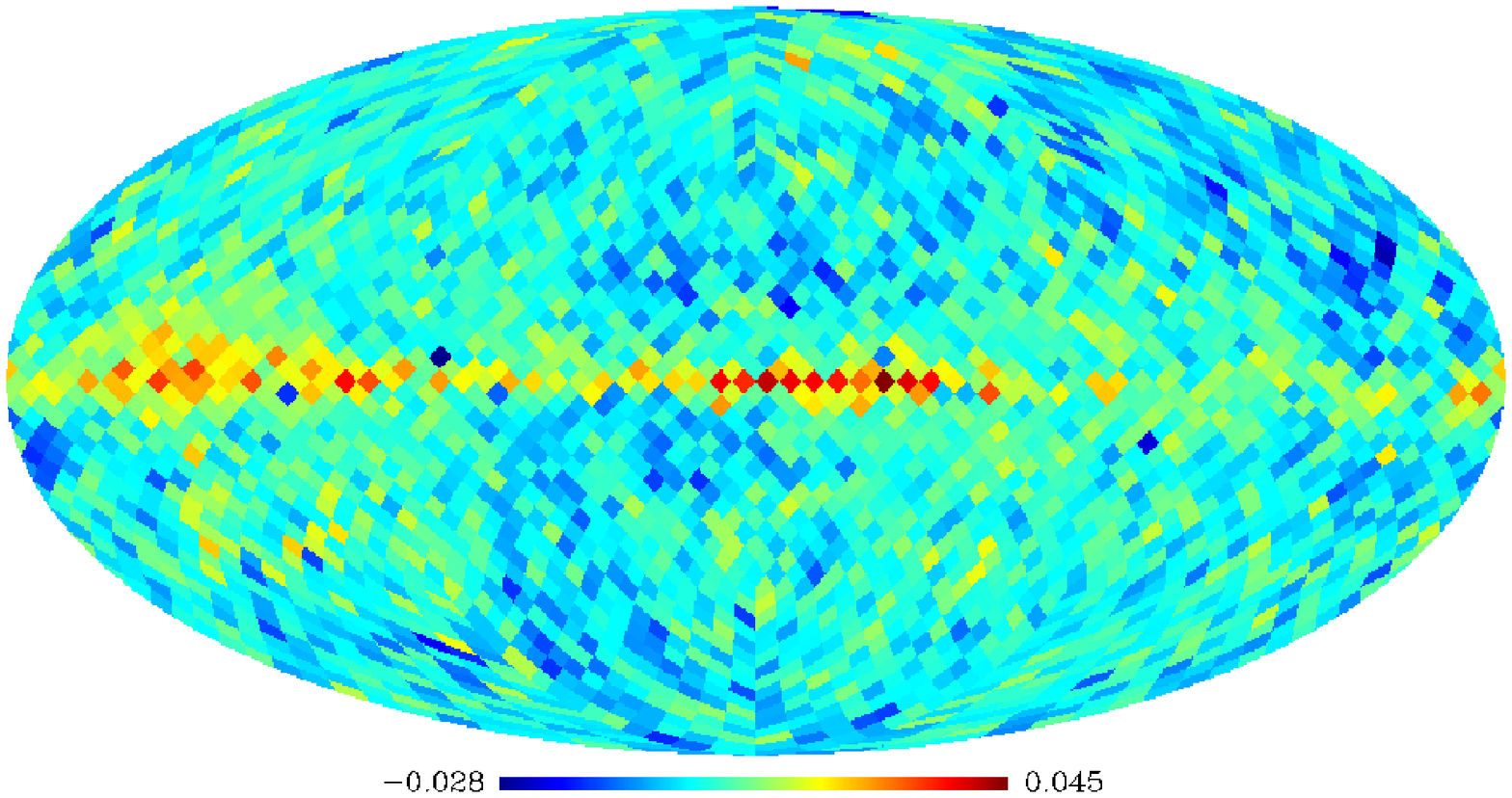}\includegraphics[angle=90,height=4cm,width=7cm]{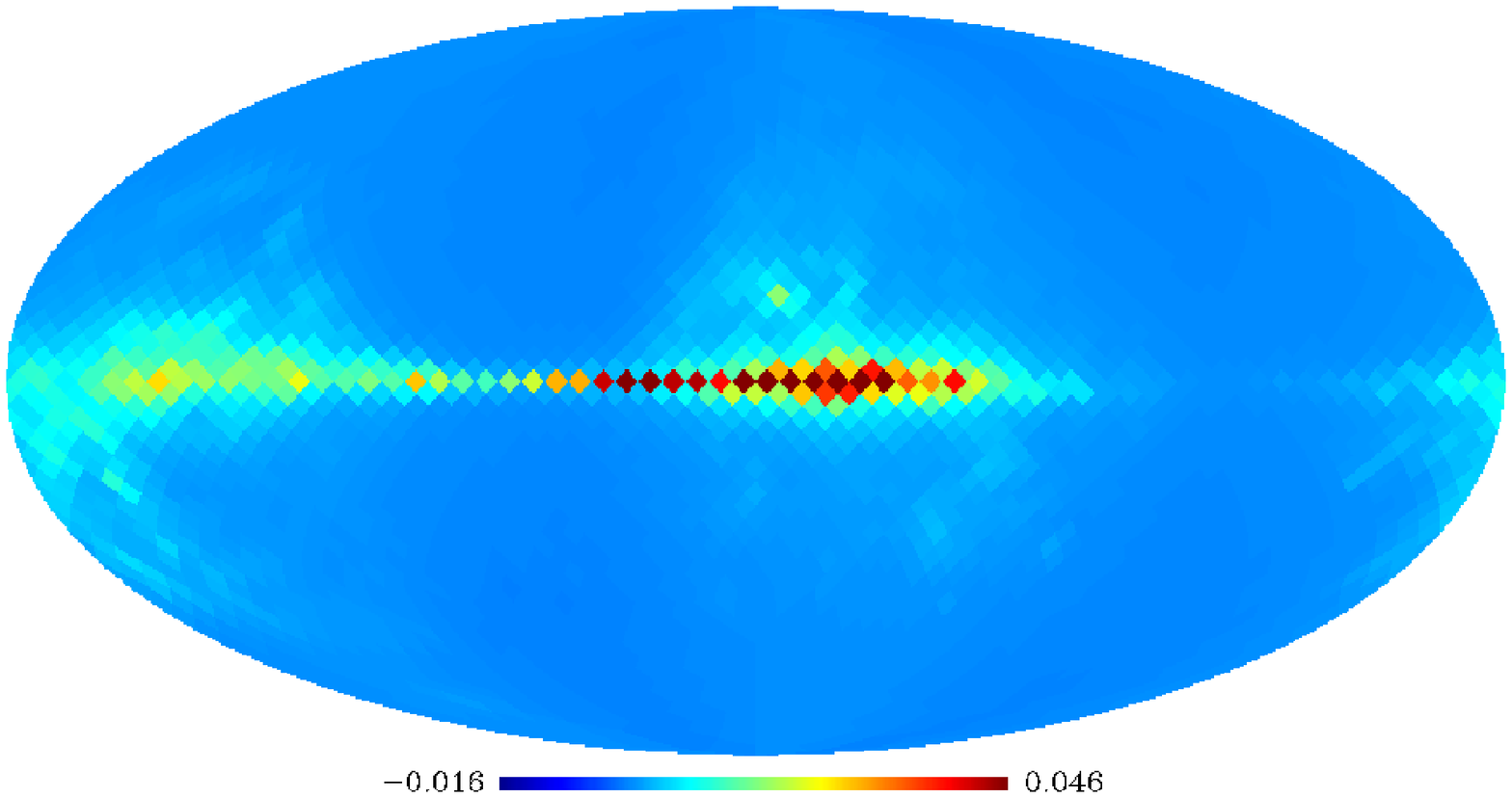}
\includegraphics[angle=90,height=4cm,width=7cm]{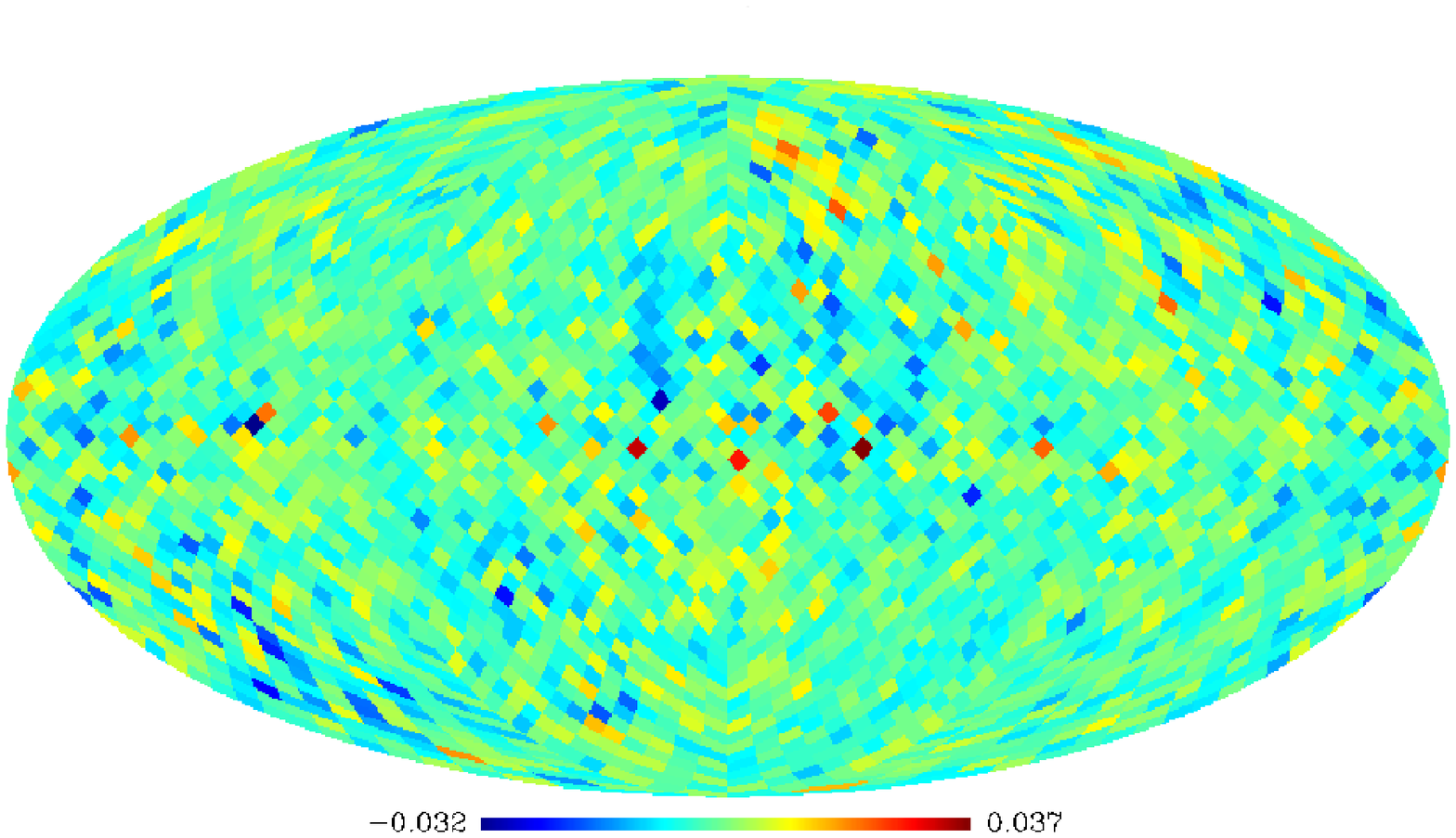}\includegraphics[angle=90,height=4cm,width=7cm]{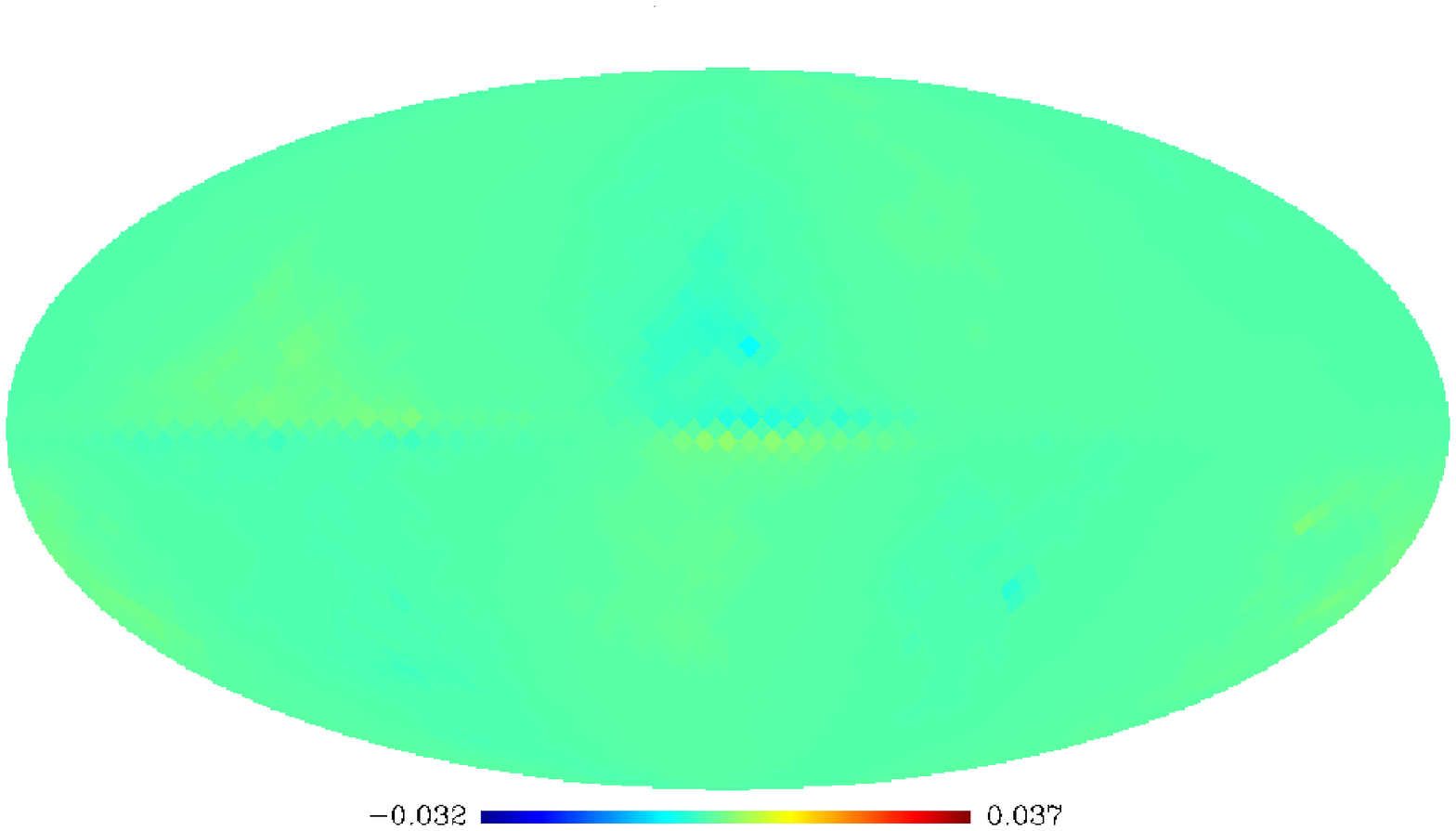}\caption{\footnotesize
  Maps (in $\mu K_{RJ}$) for the polarization Q (\emph{top}) and U  (\emph{bottom}) Stokes parameters at 94 GHz for the \Wmap\ 5 years data (\emph{left}), and for the best fit model of the polarized foreground emissions (\emph{right}) .
\label{fig:mapQUpol_rmodel94}}
\end{center}
\end{figure*}

\noindent Figures~\ref{fig:mapQUpol_rmodel33} and \ref{fig:mapQUpol_rmodel94} compare the Q (top) and U (bottom) Stokes parameter maps obtained from the best-fit parameters (right) to the \Wmap\ Ka and W band maps (left), respectively. The maps are presented in $\mu K_{RJ}$ units. At 33~GHz the synchrotron emission dominates the signal and the data 
is well represented by the model. However, at 94~GHz thermal dust emission dominates the signal. This can be observed both on the Q and U
maps, although the latter is too noisy for a clear detection of thermal dust emission.
 

\section{Conclusions}
\label{conc}

\indent We proposed in this paper an extended study of the polarized
Galactic diffuse
emissions presented in~\cite{fauvet}. We have constructed coherent
models of these two
foreground emissions based on a 3D representation of the Galactic magnetic
field and of the distributions of relativistic electrons and dust
grains in the Galaxy. For the Galactic magnetic field we considered a
 modified logarithmic spiral model for the large-scale regular
component, plus a turbulent one. The relativistic
electrons and dust grains distributions have been modeled with
exponentials peaking at the Galactic center.\\

\indent We performed a likelihood analysis to compare the available
\Wmap\ and \Archeops\ data to a set of models obtained by varying the
pitch angle of the regular magnetic field, the relative amplitude of
the turbulent magnetic field as well as the extrapolation spectral
indices for the synchrotron and thermal dust emissions.  
From this analysis, we observe that the best-fit parameters are compatible
across the frequency range explored, indicating that the
polarized sky emission at the different frequencies are of the same nature.
Using the full data set, we have been able to set constraints on the
pitch angle, $p=-30^{+15}_{-17}$ degrees. The best-fit value for the
spectral index of the synchrotron emission, $\beta_{s} = -3.45^{+0.2}_{-0.5}$,
is lower but compatible with other values found in the literature (see for example \cite{kogut,gold}). 
This low spectral index indicates that the synchrotron emission observed at microwave frequency
is produced by relativistic electrons with an steep energy spectrum (spectral index $p=4$).
 An upper limit on the relative amplitude of the turbulent component is obtained although it
seems that this turbulent part is not required to reproduce the microwave data at large angular scales.
However, we only accounted here for a statistically isotropic turbulent component and we did not for the ordered turbulent component which
in our case can not be distinguished from the regular component.\\

\indent  From above we can conclude that a simple model of the polarized Galactic
diffuse foreground emission including synchrotron and thermal dust can account
for the observed sky emission in the frequency range from 30 to 353~GHz at large
angular scales. This is of great interest for the analysis of \Planck\ satellite mission data, as \Planck\ will measure
the polarized CMB anisotropies in the frequency range from 70 to 217 GHz and we expect the
foreground emissions to be dominant at large angular scales \citep{gold2011,fauvet2012}. 

\bibliographystyle{elsarticle-harv}
\bibliography{biblio}







\end{document}